\documentclass[conference]{IEEEtran}
\IEEEoverridecommandlockouts

\usepackage{cite}
\usepackage{amsmath,amssymb,amsfonts}
\usepackage{algorithmic}
\usepackage{graphicx}
\usepackage{textcomp}
\usepackage{xcolor}
\usepackage{multirow}
\usepackage{rotating}
\usepackage{threeparttable}
\usepackage{booktabs}
\usepackage{url}

\def\BibTeX{{\rm B\kern-.05em{\sc i\kern-.025em b}\kern-.08em
    T\kern-.1667em\lower.7ex\hbox{E}\kern-.125emX}}
\begin{document}

\title{Artificial Intelligence and Civil Discourse: How LLMs Moderate Climate Change Conversations}


\author{
\IEEEauthorblockN{Wenlu Fan\IEEEauthorrefmark{1} \quad Wentao Xu\IEEEauthorrefmark{2}}
\IEEEauthorblockA{
Department of Science and Technology of Communication \\
University of Science and Technology of China, Hefei, China
}
\IEEEauthorblockA{
\IEEEauthorrefmark{2}Corresponding Author: myrainbowandsky@gmail.com 
\IEEEauthorblockA{\IEEEauthorrefmark{1} \IEEEauthorrefmark{2} These authors contributed equally to this work.}
}
}

\maketitle

\begin{abstract}
As Large Language Models (LLMs) become increasingly integrated into online platforms and digital communication spaces, their potential to influence public discourse—particularly in contentious domains like climate change—demands systematic investigation. This study examines how LLMs naturally moderate climate change conversations through their distinct communicative behaviors, offering insights into their role as facilitators of civil discourse.
We conducted a comparative analysis of conversational patterns between LLMs and human participants in climate change discussions across social media platforms. Our investigation employed five state-of-the-art models: three open-source LLMs (Gemma, Llama 3, and Llama 3.3) and two commercial systems (GPT-4o by OpenAI and Claude 3.5 by Anthropic). Through sentiment analysis, we assessed the emotional characteristics and discourse patterns exhibited by both LLMs and human users. Our findings reveal two key mechanisms through which LLMs moderate climate change conversations: First, LLMs consistently demonstrate emotional neutrality, with their responses significantly dominated by neutral sentiment compared to human participants who exhibit more polarized emotional expressions. Second, LLMs maintain notably lower emotional intensity across all interaction contexts, creating a stabilizing effect on conversational dynamics. These results suggest that LLMs possess inherent moderating capabilities that could enhance the quality of public discourse on controversial topics. By maintaining emotional equilibrium and reducing inflammatory rhetoric, LLMs may serve as valuable tools for fostering more constructive and civil climate change conversations online. This research contributes to our understanding of AI's potential role in improving digital discourse and offers implications for the design of AI-mediated communication platforms.

\end{abstract}

\begin{IEEEkeywords}
Large language model; AI cognition; Computational linguistics; Climate change; Social media; Emotion analysis; Emotion moderation; 
\end{IEEEkeywords}

\section{Introduction}
One of the most serious challenges in the 21st century that the world faces is climate change, its impacts are far-reaching, affecting the ecological environment, economic development, and the fundamental survival of human society\cite{dietz_climate_2020}. Climate change is the quintessential environmental story of our time, with potential narrative elements including big business, global economies, cutting-edge science, devastating extreme weather, and perhaps the future of civilization itself\cite{good_framing_2008}. Discussions on scientific understanding, policy making\cite{kane_linking_2000} and public action regarding climate change continue to heat up around the world, forming a complex and diverse public issue. Even though scientific research shows that climate change is happening, many people still deny it exists. The climate has changed rapidly over the past several decades\cite{masson2021climate}, this situation is expected to continue for centuries to come\cite{lyon2022climate}. Paradoxically, as climate science has become more certain, denial of the issue has increased\cite{ washington2013climate}. Powerful voices of denial have gained momentum in recent years.\cite{wong2020understanding}. These discussions not only contain the understanding of scientific fact, but also intertwined with conflicts of values, interest games, and cognitive differences among different social groups, making the public topic of climate change issues highly complex and sensitive.

In the digital time, social media platforms like Twitter (now X) and Reddit, has become key arenas for the dissemination of information, exchange of views and formation of public opinion on climate change. Social media have opened new channels for public debate, revolutionizing communication about prominent public issues such as climate change.\cite{pearce2019social}.

In the digital age, social media platforms like Twitter (now X) and Reddit have become key arenas for the dissemination of information, exchange of views, and formation of public opinion on climate change. Social media have opened up new channels for public debate and have revolutionized the communication of prominent public issues such as climate change\cite{pearce2019social}. However, these platforms’ characteristics, such as anonymity and algorithmic curation, introduce significant challenges: ‘information overload’ caused by massive amounts of information that are difficult to distinguish between true and false\cite{merlici2024too}; emotional expressions generated by users in anonymous or semi-anonymous online contexts\cite{hosseinmardi2014towards}; and "echo chambers" built by recommendation algorithms and user choices\cite{cinelli2021echo}, which increase polarization of opinions. These characteristics make it difficult to reach a consensus in climate change discussions on social media and may even solidify prejudices.

Now, the online discussion related to climate change is facing many difficulties. The spread of misinformation and disinformation not only confuses the public\cite{guess2020misinformation}, but also hinders science-based decision-making and action. Rational debate was replaced by artisanship\cite{benegal2018correcting}, the constructive dialogue was missing, making effective communication and negotiation extremely difficult. In this background, to explore how to improve the quality of network discussion and new ways to promote rational dialogue are particularly urgent. 

Recent years, artificial intelligence technology represented by Large Language Models (LLMs) has made breakthrough progress\cite{kumar2024large}. LLMs show powerful natural language understanding and generation ability by training on massive text data\cite{myers2024foundation}. They can not only process information collecting and generate fluent and natural content\cite{yao2013recurrent}, but also conduct multi-round dialogue interactions to a certain extent, simulating the way humans communicate\cite{chen2024large}.

These abilities make LLMs show broad application prospects of information service, content generation, intelligent customer service\cite{chen2024large}. Especially in social media, as an emerging communication subject, LLM has attracted academic attention for its potential to intervene in online discussions, especially when dealing with complex and controversial topics\cite{babatunde2025moderating}. Its performance and possible impact deserve further study. 

By examining these dynamics within climate change communication—a highly polarized and emotionally charged domain—this research addresses the following questions:

\textbf{\textit{RQ1: What primary emotion tendency do LLMs replies exhibit?}}

\textbf{\textit{RQ2: Compared to human text, how does the emotional expression of LLMs differ in climate change discussions?}}

By answering the research questions, this study makes the contributions as follows:
\begin{itemize}
\item{We find that LLMs exhibit a neutral emotional tendency. It will deepen the understanding of human-computer interaction, especially the behavioral patterns of AI as a communication participant in sensitive and highly controversial issues.}
\item{We systematically quantified and presented the specific emotional expression characteristics of current LLMs when dealing with real-world climate change-related online discussions. The study found that LLM replies are not only dominated by neutral sentiment, but also have significantly lower extreme emotional intensity than human texts. This lays an empirical foundation for subsequent research on the actual impact of AI on online discussions, users' perception and acceptance of AI neutral replies, and comparative analysis of the emotional expression capabilities of different LLM models.}
\item{We propose that LLM can be used as a potential tool to moderate the emotional and extremist trends in online discussions through its inherent emotional ``neutral" characteristics. This has important reference value for social media platform managers, AI system developers, and policymakers who are committed to improving the current increasingly polarized online communication ecology, helping them think about how to responsibly design and deploy LLM in order to promote healthier and more constructive public dialogues.}
\end{itemize}

\section{Related Work}
\subsection{Climate Change Discourse}
Climate change refers to long-term shifts in temperatures and weather patterns. Such shifts can be natural, due to changes in the sun’s activity or large volcanic eruptions. But since the 1800s, human activities have been the main driver of climate change, primarily due to the burning of fossil fuels like coal, oil and gas\cite{nations_what_nodate}. The surge in scientific evidence regarding climate change has elevated it to a significant societal concern\cite{corbett2022tweets, wei2021exploring}. While many people accept and engage with climate action, there are also many people hold the denialism of climate change. The 2022 report Climate Change in the American Mind finds similar results: a third of the respondents said that climate change is due to natural changes and is not caused mostly by human activities\cite{leiserowitz2019climate}. The complexity of opinion has led to many analyses of the discussion on climate change, which is also mixed with partisan opposition and business interests. Yantseva and Victoria \cite{yantseva2025climate} found that most groups promoting climate change denial occupy a common ideological space, though they tend to be located at the margins rather than the core of that space.

Samantha K. Stanley et al.\cite{stanley2024conservative} found a relationship between ideology and attitudes towards climate change that an asymmetry exists in climate-related threat perceptions, whereby greater endorsement of conservative ideology predicts lower perceived threat from climate change and greater perceived threat from climate reform.

Society values reliable information sources for constructing discourse on climate change\cite{sun_social_2024}. Lee and Theokary\cite{lee2021superstar} think that “prestige newspapers,” news aggregators or organizations, and opinion leaders, which can include celebrities, are the most trusted sources of information regarding environmental topics. Nevertheless, social networks serve as a platform for exchanging and disseminating opinions, ideas, and dialogues, transforming the 21st-century agora\cite{ ballestar2020tale, martin2022ignorance}. 

As a worldwide public topic, related to global people persistent development, considering different countries, regions and groups’ need is essential to climate change policymakers. Here, the internet is viewed as a rhetorical context providing the public with opportunities to engage with developments in science and policy, and contest elite messages\cite{koteyko_13_2015}. The decentralised and participatory nature of online social media offers a novel opportunity to study previously inaccessible aspects of social interaction about climate change\cite{auer2014potential}, including the social network structures that link individuals engaged in online debate and that are likely to affect how attitudes evolve over time.

However, social media like Twitter( now X ) also breed some negative effects. Many climate change discussions are mixed with ideology, US states with Republican voting patterns have been found to be more likely to originate Twitter comments using the term ‘global warming’ and frame it as a ‘hoax’ than were states with a preponderance of Democrat voters, where the term ‘climate change’ was more apt to be used and was framed as a real problem requiring attention\cite{ jang2015polarized}. A study of Twitter messages containing generic hashtags about climate change\cite{williams2015network} found some similar “open forums” of contestation, but concluded that discussions were more likely to take place within more homogenous enclaves of opinion. The authors conclude that ``social media discussions of climate change often occur within polarized `echo chambers.'". Such studies suggest that it is possible for online communities to contribute both to bipartisan engagement as well as enabling polarization.

While previous studies have mapped the ideological divisions and information dynamics of climate change discussions, they largely focus on human-generated content and overlook the increasing role of AI-generated text. Little is known about how LLMs, when engaged in these discussions, may adopt or diverge from the emotional patterns typically observed in human discourse. This gap is especially relevant as LLMs are increasingly deployed on public platforms and may influence online climate narratives.

\subsection{Online Deliberation and Polarization}
Nowadays, the polarization has been observed over numerous societal, political and economic issues\cite{einav_bursting_2022}. Media plays a considerable role. Online users could selectively served information based on their own personal information. Courtois et al.\ \cite{courtois2018challenging} argue that such an environment fosters various cognitive biases, particularly confirmation bias, which refers to the tendency to seek or interpret information in a way that aligns with pre-existing beliefs, expectations, or hypotheses. The existence of “filter bubble” and “echo chamber” hinders pluralistic dialogue and thus jeopardizes democracies in modern society\cite{bozdag2015breaking}. These consequences lead to the generation of, living within a filter bubble and exposure to racist, sexist, or homophobic views may lead to desensitization and further spread of discriminatory materials\cite{kilvington2021virtual}. Furthermore, polarization in its extreme form may also lead to social/national fragmentation and even civil war\cite{harel2020conflict}.

Beyond the traditional opinion polarization, more studies are turning to the other aspects of polarization. Iyengar et al.\cite{iyengar2019origins} proposed a that a new type of division has emerged in the mass public in recent years. Ordinary Americans increasingly dislike and distrust those from the other party. This phenomenon of animosity between the parties is known as affective polarization. Lerman et al.\cite{Lerman2024} measure affective polarization on social media by quantifying the emotions and toxicity of reply interactions. And they found that interactions between users with same ideology (in-group replies) tend to be positive, while interactions between opposite-ideology users (out-group replies) are characterized by negativity and toxicity. Simchon et al.\cite{simchon_troll_2022} focused on the language of online polarization and examined the processes by which trolls may sow intergroup conflict through polarized rhetoric.

About the polarization reducing, many researches are focusing on the affective polarization of party ideology. Rossiter and Carlson\cite{Rossiter2024} found that cross-partisan conversations reduced out-party animosity for at least three days, reduced social polarization.
Ahler and Sood \cite{ahler2018parties} show that Americans tend to hold exaggerated stereotypes about the demographic composition of opposing parties—for example, overestimating the number of union members among Democrats or high-income individuals among Republicans—leading to inflated perceptions of ideological distance. Their study finds that correcting these misperceptions reduces partisan hostility, as individuals come to view out-party members as less different and more relatable. Levendusky\cite{levendusky2016mis} further demonstrates that shifting focus from partisan identity to a shared national identity—emphasizing that Democrats and Republicans are both Americans—can significantly reduce affective polarization. Participants primed with national identity were more likely to rate members of the opposing party more favorably, suggesting that re-framing group membership can mitigate intergroup animus. 

Moreover, while survey experiments provide promising evidence, their external validity remains uncertain, and more work is needed to test whether these effects generalize to real-world political contexts. Emerging directions draw on intergroup contact theory\cite{pettigrew2013groups} and the role of cross-cutting social networks\cite{mutz2002cross}, suggesting that structured dialogue between ordinary partisans could reduce hostility, though the scalability and long-term impact of such efforts remain to be evaluated.

Most existing research focuses on human interactions and the psychological or structural mechanisms of polarization, particularly affective polarization. However, the potential moderating role of non-human agents—such as LLMs—in these conversations remains underexplored. Can LLMs amplify or reduce polarization through their emotional tone? Do their responses tend toward neutrality or reflect existing biases? These are important yet unanswered questions in the current literature.

\subsection{LLMs in Social Media}
Large Language Models (LLMs), such as ChatGPT and Bard, have revolutionized natural language understanding and generation\cite{YAO2024100211}. A capable LLM should exhibit four key features\cite{yang2024harnessing}: (i) profound comprehension of natural language context; (ii) ability to generate human-like text; (iii) contextual awareness, especially in knowledge-intensive domains; (iv) strong instruction-following ability which is useful for problem-solving and decision-making. As their extraordinary ability is growing, LLMs are increasingly being developed as proactive communication partners. Argyle et al.\cite{doi:10.1073/pnas.2311627120} think that prompted LLMs can effectively roleplay believable characters. Agentbased simulations, powered by LLMs, have been used for understanding debates\cite{du2023improvingfactualityreasoninglanguage},strategic communication\cite{ xu2024exploringlargelanguagemodels }, conflict\cite{hua2024warpeacewaragentlarge} and online behavior\cite{ ren2024baseslargescalewebsearch}. These studies mark an important shift in the role of the LLM, from a passive tool to an interactive agent in the communication process. 
Another application scenario of LLM is social media. LLMs has revolutionized content generation, offering sophisticated tools for text generation, dialogue systems, and multimedia content creation\cite{thapa2025large}. And chatbots based on LLMs are now more accurate, creative, context-aware, and human-like in interactions\cite{naik2023demystifying}. AI chatbots are able to assist individuals, allowing them to manage and prioritise tasks\cite{naik2024applications}. Nevertheless, the ability of language models to communicate with humans is no longer limited to understanding instructions. There is study\cite{ chun2025conflictlens} have shown that large language models can also be used to help humans deal with deep psychological problems. 

Despite the rapid development of LLMs and their integration into social media platforms, few studies have empirically investigated the emotional characteristics of their generated content in highly polarized contexts like climate change. While some works explore LLMs' capabilities for role-play or emotional support, their actual emotional stance, neutrality, or possible bias in contentious public discussions is not well understood. Moreover, their potential role in moderating polarized discussions—by reducing affective intensity or introducing neutrality—remains largely speculative. Our study aims to fill this gap by systematically comparing the emotional tendencies of LLMs with human-generated texts in climate-related discourse.

In sum, while existing literature has advanced our understanding of climate change discourse, online polarization, and the capabilities of LLMs, an important problem remains: how do LLMs behave emotionally in polarized public discourse, and do they exhibit a moderating effect through emotional neutrality? This study addresses this problem by focusing on LLM-generated responses to climate change discussions and comparing their emotional expression with that of human participants.

\section{Methodology}
\subsection{Experiment}
\begin{figure*}[htb]
\centering 
\includegraphics[width=0.6\linewidth, height=0.4\textwidth]{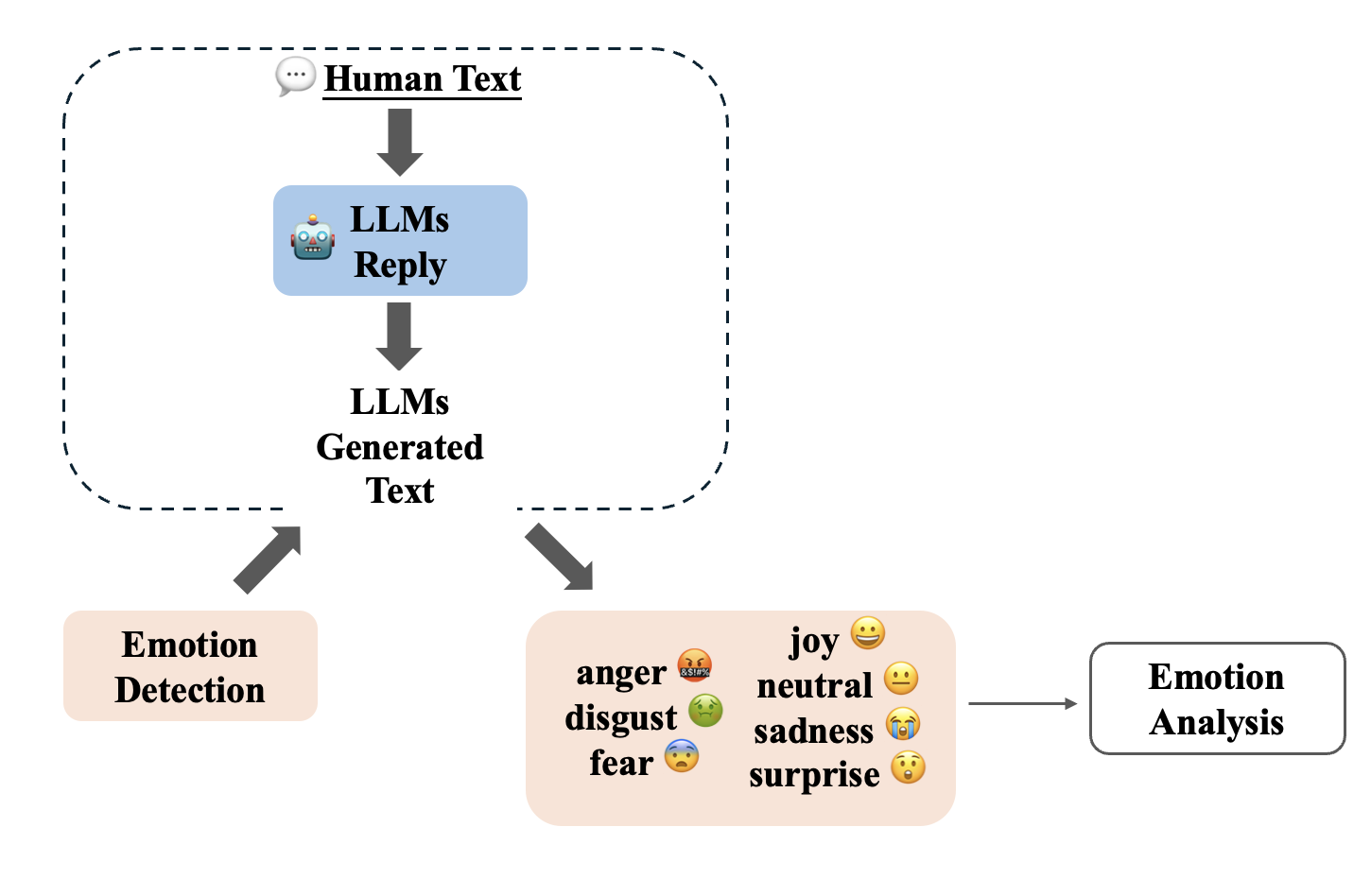}
\caption{Experimental pipeline of emotion tendency for LLMs. Our experimental framework begins with human text input to five LLMs , which perform reply tasks. Following content generation, we implement emotion detection on the outputs, followed by comprehensive analyses. The framework concludes with parallel analyses of emotional tendency and emotional intensity to evaluate the emotional pattern of LLM-generated content relative to the original human input.}
\label{Experiment design}
\end{figure*}

Figure~\ref{Experiment design} presents the structure of our experimental setup. The central aim of our study is to explore how large language models (LLMs) respond to human-generated content, specifically social media posts sourced from real users.

We employed three open-source LLMs—Gemma\footnote{\url{https://ai.google.dev/gemma}} (developed by Google), Llama 3\footnote{\url{https://ai.meta.com/llama/license/}}, and Llama 3.3 (both developed by Meta)—alongside two commercial models: GPT-4o by OpenAI\footnote{\url{https://openai.com}} and Claude 3.5 by Anthropic\footnote{\url{https://www.anthropic.com/claude}}. The specific model versions used include Gemma2-27B-Instruct-Q8, Llama3-70B-Instruct, Llama3.3-70B, GPT-4o-all, and Claude-3-5-Haiku-20241022-X. These models were chosen for their demonstrated strength and reliability in language understanding and generation. To run the open-source models locally, we utilized the Ollama framework\footnote{https://ollama.com/}.

For the reply generation task, we had each model reply to individual human-authored posts through a chat function. This setup allowed the models to engage with the input naturally, without requiring manually crafted instruction prompts, thereby simulating a realistic conversational context. We clarify that the original user-generated post served as the prompt for the model’s reply, without additional explicit instructions.

\subsection{Dataset} 
This study utilized climate change corpora collected from Twitter (now X) and Reddit. We collected data using the Twitter Search API by querying relevant keywords, including ``climate change'', ``climate science'', ``climate manipulation'', ``climate Engineering'', ``climate Hacking'', ``climate modification'', ``Global Warming'', ``carbon footprint'', and ``The Paris Agreement''. For Reddit, we used data maintained by Pushshift from https://the-eye.eu/redarcs/. The Pushshift Reddit dataset consists of two sets of files: submissions and comments ~\cite{DBLP:journals/corr/abs-2001-08435}. The same keywords were applied to filter Reddit data, and to compare the differences in emotions, we collected both posts and comments from both platforms.

\begin{figure*}[ht]
\begin{minipage}[ht]{0.5\linewidth}
\centering
\includegraphics[width=0.7\textwidth, height=0.5\textwidth]{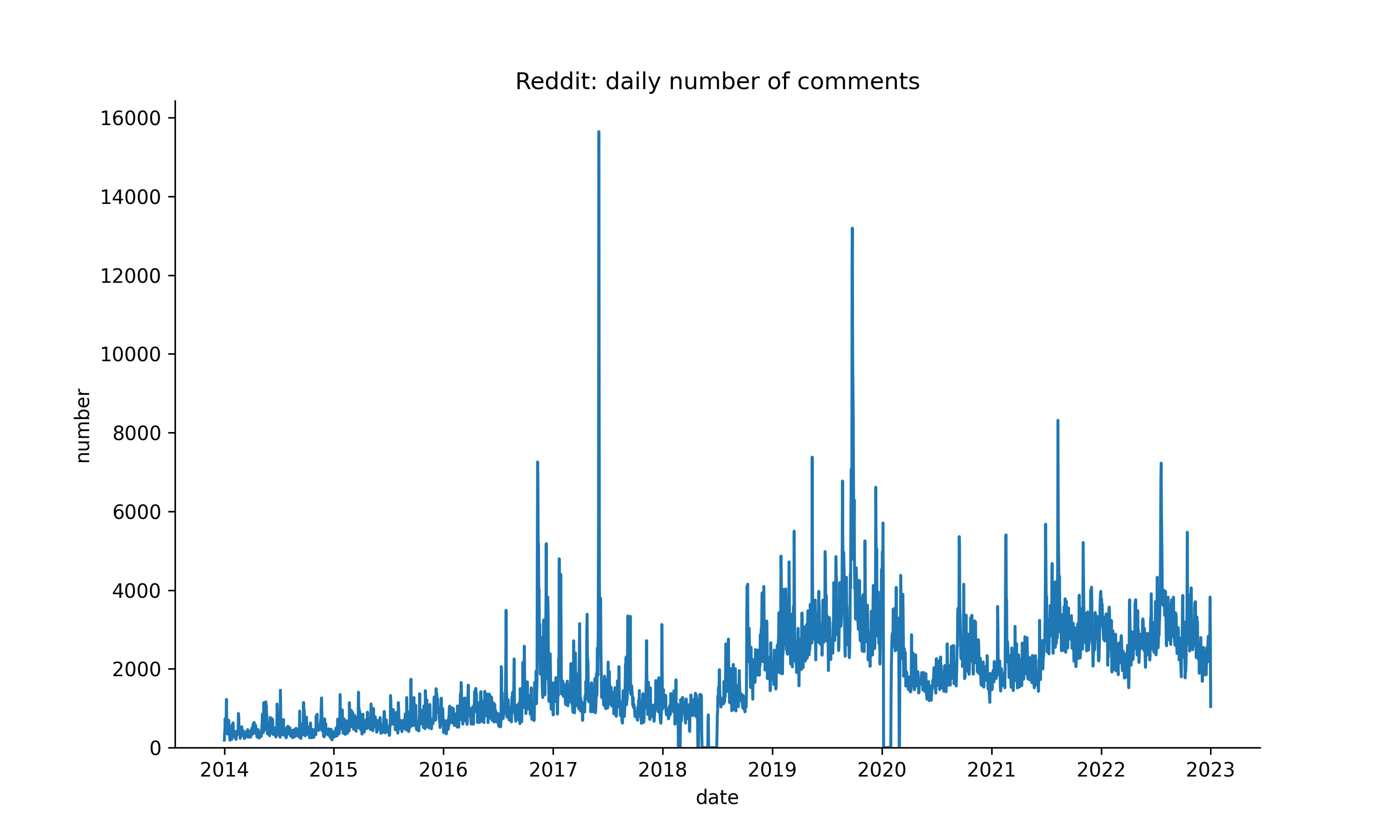}
\centerline{\textbf{a}}
\end{minipage}%
\begin{minipage}[ht]{0.5\linewidth}
\centering
\includegraphics[width=0.7\textwidth, height=0.5\textwidth]{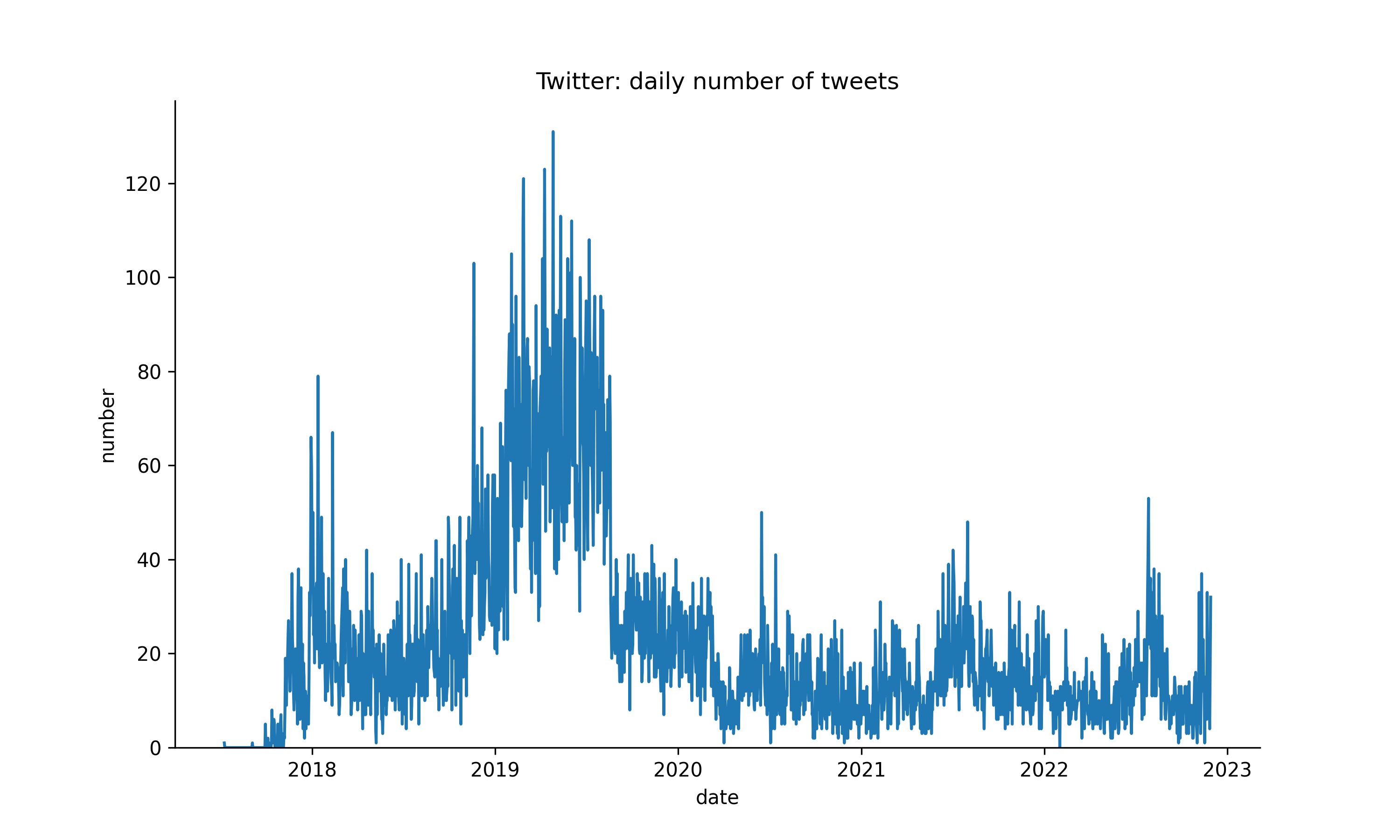}
\centerline{\textbf{b}}
\end{minipage}
\caption{Daily data amount of Twitter and Reddit.
\textbf{a.} Daily comments count of Reddit. \textbf{b.} Daily tweets count of Twitter. The x-axis represents the date, and the y-axis represents the frequency.}
\label{daily distribution}
\end{figure*}

With the keywords, we obtained 5,768,822 Reddit comments and 76,596,654 tweets from Twitter. We used histograms to understand the basic distribution of data (Figure~\ref{daily distribution}). 

Twitter and Reddit serve as two prominent yet distinct social media platforms. Twitter emphasizes brevity and immediacy, encouraging rapid, real-time communication through short messages. In contrast, Reddit is structured around topic-specific communities that support more in-depth, reflective conversations\cite{10216752, Treen04072022}. These contrasting platform architectures influence not only the nature of user interactions and content sharing but also appeal to different types of users with varying interests and participation patterns. As a result, they offer complementary perspectives for analyzing diverse forms of public discourse\cite{ruan2022cross}.

To develop datasets that accurately reflect the long-term discourse patterns related to climate change while minimizing bias from sudden, event-driven spikes, we adopted a time-stratified sampling strategy. Drawing from a raw corpus of 5,768,822 Reddit comments and 76,596,654 Twitter posts, we first segmented the data by month. Within each monthly subset, we then applied systematic sampling—selecting 200 entries from Twitter and 100 from Reddit. This month-by-month proportional sampling approach ensured even temporal coverage, helping to reduce distortion from short-lived surges in activity or interest. The goal was to build a dataset that captures both the substance and progression of climate change conversations over time. This method ultimately produced a refined dataset of 12,200 Twitter samples and 10,900 Reddit samples for analysis.

\subsection{Emotion Labeling}

In this study, we propose a method for analyzing emotions in cross-platform social media data using a deep neural network-based approach. Instead of traditional emotion analysis, which classifies text as positive, negative, or neutral, we adopt a fine-grained, emotion-oriented framework to better capture the nuanced affective states present in both human-written and LLM-generated content\cite{guo2023close}.

To implement this approach, we utilized the \textit{j-hartmann/emotion-english-distilroberta-base}\footnote{https://huggingface.co/j-hartmann/emotion-english-distilroberta-base} model from Hugging Face. This model is a fine-tuned checkpoint of “DistilRoBERTa-base,” built upon the RoBERTa-base architecture, and trained on datasets annotated with emotional labels across diverse sources, including Twitter, Reddit, student self-reports, and television dialogues.

The model classifies text into seven distinct emotion categories: joy, surprise, neutral, anger, disgust, fear, and sadness. It provides probability scores for each category, which serve as quantitative indicators of emotional expression and form the basis for our subsequent analyses.

\section{Result}
\subsection{Neutral Emotion Predominates in LLM Replies}
In this study, we examined the emotional transitions between human-generated text and LLM outputs in downstream tasks. We categorize 7 emotions as positive emotionally oriented and negatively oriented as well as neutral emotions as defined by itself, as follows~\cite{robinson2008brain} :

Positive emotions: anticipation, joy, love, optimism, surprise, trust ~\cite{vaillant2008positive};

Negative emotions: anger, disgust, fear, pessimism, sadness

\begin{figure*}[ht]
\centering
\includegraphics[width=1\linewidth]{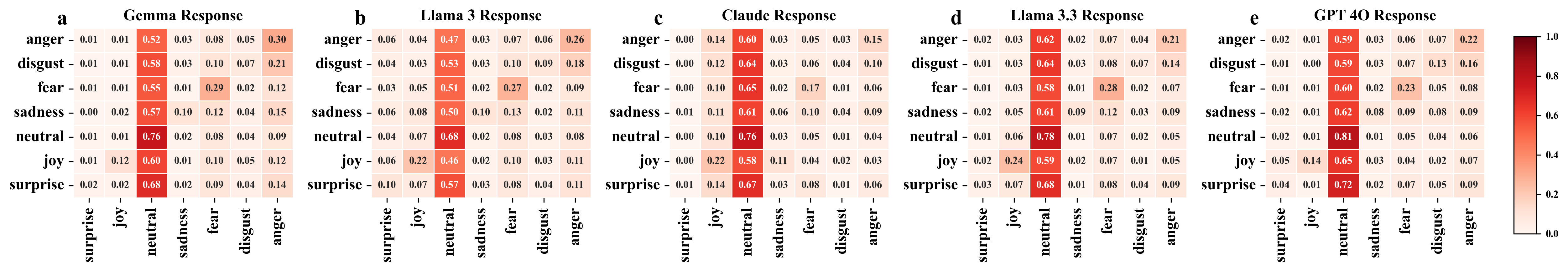}
\caption{Emotional Transition of LLM Replies in Reddit Comments.
Panels \textbf{a}, \textbf{b}, \textbf{c}, \textbf{d} and \textbf{e} illustrate emotional transitions in replies generated by Gemma, Llama, Claude and GPT models, respectively. The y-axis represents source emotions from human text, while the x-axis indicates emotions in LLM-generated content. Values labled on each cell represent the proportion of emotional transitions between original and generated content. For example, in Figure \textbf{3a}, the value 0.34 in the anger-to-anger cell indicates that 30\% of originally angry texts maintained their emotional valence in Gemma's reply task. The intensity of each cell's shading represents the proportion of emotional transition, with darker shades indicating higher transition frequencies.}
\label{Reddit:heatmap}
\end{figure*}

From Figure~\ref{Reddit:heatmap}, we can see the transition of emotion of LLMs' replies. Figure~\ref{Reddit:heatmap}\textbf{a} shows that in the Gemma' conversation, 52\% anger original texts have the neutral replies. For other emotion categories, including 'disgust,' 'sadness,' 'neutral,' 'joy,' and 'surprise,' the proportions of texts that turned into angry emotional label were 58\%, 55\%, 57\%, 60\%, and 68\%, respectively. We observed the similar result in the other LLMs' replies, demonstrating a neutral emotion moderation.

The difference between the open-source and commercial models is not significant, as they all exhibited a neutral emotional tendency in their replies. Nevertheless, we found that even within the same model family, there can be notable differences. A comparison of Figure~\ref{Reddit:heatmap}\textbf{b} and Figure~\ref{Reddit:heatmap}\textbf{d} shows that Llama 3.3 produced a higher proportion of neutral replies than Llama 3. This may be because Llama 3.3 has larger training parameters and newer experimental data.

\begin{figure*}[ht]
\centering
\includegraphics[width=1\linewidth]{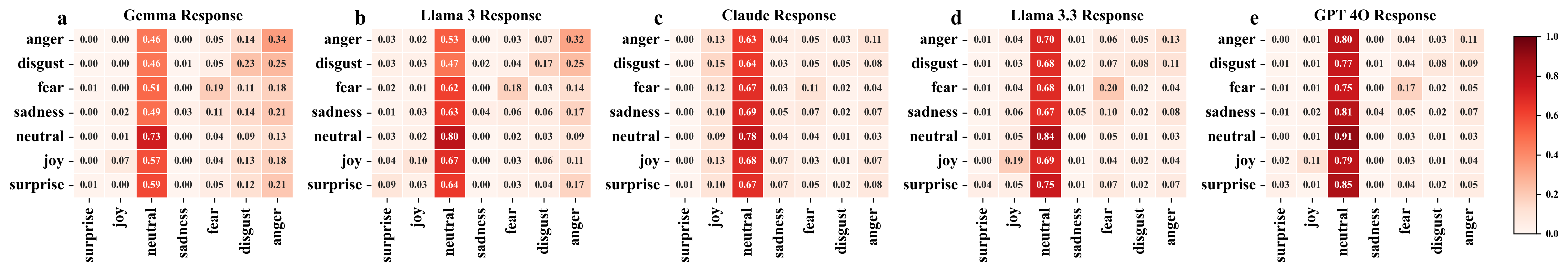}
\caption{Emotional Transition of LLM Replies in Twitter Comments.
Panels \textbf{a}, \textbf{b}, \textbf{c}, \textbf{d} and \textbf{e} illustrate emotional transitions in replies generated by Gemma, Llama, Claude and GPT models, respectively. The y-axis represents source emotions from human text, while the x-axis indicates emotions in LLM-generated content. Values labled on each cell represent the proportion of emotional transitions between original and generated content.}
\label{Twitter:heatmap}
\end{figure*}

Except Reddit, we also analyzed data from Twitter. Figure~\ref{Twitter:heatmap} illustrates the emotional transitions between original Twitter posts and LLM-generated replies, revealing similar discourse patterns surrounding climate change. On Twitter, when responding to human-authored content, LLMs likewise demonstrate a tendency toward emotional moderation, particularly by exhibiting neutral emotional replies.
In addition, we found that Claude's replies tend to exhibit a certain degree of positive emotion, which is less apparent in the outputs of other models. This suggests that Claude is not only neutral in tone, but also shows a comparatively greater inclination toward positive emotional expression.

\subsection{Lower Emotional Intensity of LLM replies}
We compared  LLM-generated content's emotional intensity with human-authored content, focusing on whether LLMs exhibit higher or lower emotional intensity. Emotional content was quantified using a probabilistic model assigning normalized scores (0 to 1) to each emotional category, interpreted as intensity scores \cite{miyazaki2024impact}. We employed ANOVA test to compare group differences and Tukey’s post-hoc test for significant pairwise variations\cite{article}. 

Human text and text generated by five models constitute six groups of data with different emotion scores. Anova test is performed on multiple groups of data to test whether there are significant differences in different sentiment scores for the same sentiment label.
As shown in Table \ref{ANOVA Test}, ANOVA revealed statistically significant differences ($P < 0.01$) in the intensity of seven emotions—anger, disgust, fear, sadness, joy, surprise, and neutral—across nine groups on Twitter. Similar emotional variation patterns were observed in the Reddit data. Post-hoc analysis using Tukey's test further identified numerous significant pairwise differences, notably in both inter-model comparisons and comparisons between model-generated and human-generated content.

Both LLM- and human-generated texts were annotated with corresponding emotion intensity scores, where higher scores indicate stronger emotional expression. When comparing across models, we observed that LLM-generated texts generally exhibit significantly lower emotion scores than human-authored texts for emotions such as joy, anger, and neutrality. Notably, certain platform-specific patterns emerged. On Twitter, fear intensity was markedly higher in human-authored texts than in LLM replies, whereas on Reddit, this discrepancy was most pronounced in the case of disgust.

These findings yield two important insights. First, LLMs tend to express emotions with lower intensity, which may reflect a more subdued or moderated emotional style. Second, the observed cross-platform differences suggest that LLM responses are influenced by the specific context and content structure of each social media platform. Together, these results highlight the necessity of accounting for platform-specific dynamics when assessing the emotional fidelity of LLM-generated text.

\begin{table}[htbp]
\centering
\begin{threeparttable}
\caption{ANOVA Test of Different Model Emotions}
\label{ANOVA Test}

\begin{tabular}{cccc}
\toprule
Platform & Emotions\tnote{a} & F Statistic & P Value\tnote{b} \\
\midrule

\multirow{7}{*}{Reddit}
& anger     & 203.0364 & $<0.001$ \\
& joy       & 50.2393  & $<0.001$ \\
& disgust   & 42.9026  & $<0.001$ \\
& surprise  & 31.0947  & $<0.001$ \\
& fear      & 0.4045   & $>0.05$  \\
& sadness   & 4.2491   & $<0.001$ \\
& neutral   & 294.4816 & $<0.001$ \\
\midrule

\multirow{7}{*}{Twitter}
& anger     & 170.0542 & $<0.001$ \\
& joy       & 55.1242  & $<0.001$ \\
& disgust   & 31.1522  & $<0.001$ \\
& surprise  & 61.9039  & $<0.001$ \\
& fear      & 21.8768  & $<0.001$ \\
& sadness   & 3.2915   & $<0.001$ \\
& neutral   & 744.5498 & $<0.001$ \\
\bottomrule
\end{tabular}

\begin{tablenotes}
\footnotesize
\item[a] Emotion scores derived from five LLM-generated contents and human texts. The five LLMs are Gemma, Llama 3, Llama 3.3, GPT-4o, and Claude 3.5.
\item[b] The $p$ values indicate the significance level of the ANOVA test; $p<0.001$ suggests statistically significant differences.
\end{tablenotes}
\end{threeparttable}
\end{table}

\begin{sidewaystable}[htbp]
\caption{Tukey's post-hoc test of LLM-generated content emotion values on Reddit}
\label{comparison_models}
\centering
\begin{threeparttable}
\resizebox{\textwidth}{!}{%
\begin{tabular}{cccccccccccccccc}
\toprule
         & \multicolumn{10}{c}{\textbf{Between models\tnote{a}}} & \multicolumn{5}{c}{\textbf{Compared with Human Texts\tnote{b}}} \\
\cmidrule(lr){2-11} \cmidrule(lr){12-16}
\textbf{Emotions} & \textbf{\begin{tabular}[c]{@{}c@{}}Gemma\\ vs Llama\end{tabular}} 
& \textbf{\begin{tabular}[c]{@{}c@{}}Gemma\\ vs Claude\end{tabular}} 
& \textbf{\begin{tabular}[c]{@{}c@{}}Gemma\\ vs Llama3.3\end{tabular}} 
& \textbf{\begin{tabular}[c]{@{}c@{}}Gemma\\ vs GPT-4o\end{tabular}} 
& \textbf{\begin{tabular}[c]{@{}c@{}}Llama\\ vs Claude\end{tabular}} 
& \textbf{\begin{tabular}[c]{@{}c@{}}Llama\\ vs Llama3.3\end{tabular}} 
& \textbf{\begin{tabular}[c]{@{}c@{}}Llama\\ vs GPT-4o\end{tabular}} 
& \textbf{\begin{tabular}[c]{@{}c@{}}Claude\\ vs Llama3.3\end{tabular}} 
& \textbf{\begin{tabular}[c]{@{}c@{}}Claude\\ vs GPT-4o\end{tabular}} 
& \textbf{\begin{tabular}[c]{@{}c@{}}Llama3.3\\ vs GPT-4o\end{tabular}} 
& \textbf{\begin{tabular}[c]{@{}c@{}}Gemma\\ vs Original\end{tabular}} 
& \textbf{\begin{tabular}[c]{@{}c@{}}Llama\\ vs Original\end{tabular}} 
& \textbf{\begin{tabular}[c]{@{}c@{}}Claude\\ vs Original\end{tabular}} 
& \textbf{\begin{tabular}[c]{@{}c@{}}Llama3.3\\ vs Original\end{tabular}} 
& \textbf{\begin{tabular}[c]{@{}c@{}}GPT-4o\\ vs Original\end{tabular}} \\
\midrule
Disgust  & - & - & - & - & - & - & - & - & - & - & \( > \) *** & \( > \) *** & \( > \) *** & \( > \) *** & \( > \) *** \\
Fear     & - & - & - & - & - & - & - & - & - & - & - & - & - & - & - \\
Anger    & \( < \) *** & \( < \) * & - & - & \( < \) *** & \( > \) *** & - & - & \( < \) *** & - & \( < \) *** & \( < \) *** & \( < \) *** & \( < \) *** & \( < \) *** \\
Surprise & \( > \) *** & \( < \) * & \( > \) * & - & - & - & \( > \) *** & - & - & - & - & \( < \) *** & - & - & - \\
Joy      & \( > \) * & \( < \) ** & \( > \) *** & - & - & \( > \) *** & \( > \) * & \( > \) *** & \( < \) ** & \( > \) *** & - & \( < \) * & \( < \) ** & \( < \) *** & - \\
Neutral  & \( < \) *** & \( > \) *** & \( > \) * & - & \( < \) *** & \( > \) *** & \( < \) *** & \( > \) *** & \( > \) *** & - & \( < \) *** & \( < \) *** & \( < \) *** & \( < \) *** & \( < \) *** \\
Sadness  & \( < \) ** & \( > \) *** & - & - & \( > \) *** & - & - & \( > \) *** & \( > \) *** & - & - & - & \( > \) *** & - & - \\
\bottomrule
\end{tabular}
}
\begin{tablenotes}
\footnotesize
\item[a] Comparison of emotion values from content generated by different models.
\item[b] Comparison between model-generated content and original human text.
\item[c] Stars indicate the \textit{p}-values of the Mann–Whitney U test: *** for \( p<0.001 \), ** for \( p<0.01 \), and * for \( p<0.05 \).
\item[d] The symbols \( < \) and \( > \) indicate that the value in the previous column is significantly less than or greater than that in the next column. A “-” indicates no significant difference.
\end{tablenotes}
\end{threeparttable}
\end{sidewaystable}

\begin{sidewaystable}[htbp]
\caption{Tukey's post-hoc test of LLM-generated content emotion values on Twitter}
\label{comparison_models_twitter}
\centering
\begin{threeparttable}
\resizebox{\textwidth}{!}{%
\begin{tabular}{cccccccccccccccc}
\toprule
         & \multicolumn{10}{c}{\textbf{Between models\tnote{1}}} & \multicolumn{5}{c}{\textbf{Compared with Human Texts\tnote{1}}} \\
\cmidrule(lr){2-11} \cmidrule(lr){12-16}
\textbf{Emotions} & \textbf{\begin{tabular}[c]{@{}c@{}}Gemma\\ vs Llama\end{tabular}} 
& \textbf{\begin{tabular}[c]{@{}c@{}}Gemma\\ vs Claude\end{tabular}} 
& \textbf{\begin{tabular}[c]{@{}c@{}}Gemma\\ vs Llama3.3\end{tabular}} 
& \textbf{\begin{tabular}[c]{@{}c@{}}Gemma\\ vs GPT-4o\end{tabular}} 
& \textbf{\begin{tabular}[c]{@{}c@{}}Llama\\ vs Claude\end{tabular}} 
& \textbf{\begin{tabular}[c]{@{}c@{}}Llama\\ vs Llama3.3\end{tabular}} 
& \textbf{\begin{tabular}[c]{@{}c@{}}Llama\\ vs GPT-4o\end{tabular}} 
& \textbf{\begin{tabular}[c]{@{}c@{}}Claude\\ vs Llama3.3\end{tabular}} 
& \textbf{\begin{tabular}[c]{@{}c@{}}Claude\\ vs GPT-4o\end{tabular}} 
& \textbf{\begin{tabular}[c]{@{}c@{}}Llama3.3\\ vs GPT-4o\end{tabular}} 
& \textbf{\begin{tabular}[c]{@{}c@{}}Gemma\\ vs Original\end{tabular}} 
& \textbf{\begin{tabular}[c]{@{}c@{}}Llama\\ vs Original\end{tabular}} 
& \textbf{\begin{tabular}[c]{@{}c@{}}Claude\\ vs Original\end{tabular}} 
& \textbf{\begin{tabular}[c]{@{}c@{}}Llama3.3\\ vs Original\end{tabular}} 
& \textbf{\begin{tabular}[c]{@{}c@{}}GPT-4o\\ vs Original\end{tabular}} \\
\midrule
Joy      & \( > \) ***  & -        & \( > \) *** & -        & \( < \) *** & \( > \) *** & \( < \) *   & \( > \) *** & \( > \) ** & \( < \) *** & -        & \( < \) *** & -        & \( < \) *** & -        \\
Neutral  & \( > \) ***  & \( > \) *** & \( > \) *** & \( > \) *** & \( < \) *** & \( > \) *** & \( > \) *** & \( > \) *** & \( > \) *** & \( > \) *** & \( < \) *** & \( < \) *** & \( < \) *** & \( < \) *** & \( < \) *** \\
Fear     & -           & -        & -           & -        & -           & -           & -           & -           & -           & -           & \( > \) *** & \( > \) *** & \( > \) *** & \( > \) *** & \( > \) *** \\
Surprise & \( > \) *    & -        & -           & -        & -           & \( < \) *** & \( < \) *** & -           & -           & -           & -        & \( < \) *** & -        & \( < \) **  & -        \\
Sadness  & -           & \( < \) *** & -           & -        & \( < \) **  & -           & -           & \( > \) *** & \( > \) *** & -           & -        & -        & \( > \) *** & \( < \) *   & -        \\
Anger    & \( > \) ***  & \( > \) *** & \( > \) *** & \( > \) *  & \( > \) **  & \( > \) *** & \( < \) *** & -           & \( < \) *** & \( < \) *** & \( < \) *** & \( < \) *** & \( < \) *** & \( < \) *** & \( < \) *** \\
Disgust  & -           & -        & -           & -        & -           & -           & -           & -           & -           & -           & \( > \) *** & \( > \) *** & \( > \) *** & \( > \) *** & \( > \) *** \\
\bottomrule
\end{tabular}
}
\begin{tablenotes}
\footnotesize
\item[a] Comparison of emotion values from content generated by different models.
\item[b] Comparison between model-generated content and original human text.
\item[c] Stars indicate the \textit{p}-values of the Mann–Whitney U test: *** for \( p<0.001 \), ** for \( p<0.01 \), and * for \( p<0.05 \).
\item[d] The symbols \( < \) and \( > \) indicate that the value in the previous column is significantly less than or greater than that in the next column. A “-” indicates no significant difference.
\end{tablenotes}
\end{threeparttable}
\end{sidewaystable}

\section{Discussion}

This study aims to analyze the emotional characteristics of LLM responses to climate change-related statements from social media users and investigate the potential role of these characteristics in moderating discussions.

\subsection{Neutral Emotion}
Regarding RQ1, our findings indicate that LLM-generated replies predominantly exhibit emotional neutrality. That is, when prompted with climate change-related user statements, LLMs tend to respond with text that lacks strong emotional valence—neither overtly positive nor negative. This consistent neutrality emerges as the most salient emotional characteristic of LLMs in this context. Rather than amplifying emotional tones, LLMs maintain a calm, measured response style, suggesting an intrinsic tendency to remain emotionally detached when engaging with sensitive or controversial topics such as climate change.

\subsection{Low Emotion Intensity}
With respect to RQ2, this study reveals a significant attenuation of emotional intensity in LLM replies compared to human-generated content. Human texts on climate change often display high emotional salience—such as anger, fear, or urgency—reflecting the deeply personal and politically charged nature of the topic. In contrast, LLM responses not only avoid strong emotional expression but also exhibit a markedly lower emotional "concentration." Even when reacting to emotionally charged user inputs, LLMs consistently respond with reduced intensity, often shifting the tone toward neutrality or composure. This contrast suggests that LLMs have a natural dampening effect on emotionally heated discourse.

Together, these two findings highlight a distinctive and potentially valuable function of LLMs in online discussions: their inherent capacity for emotional cooling or emotional debiasing. By resisting emotional provocation and maintaining low-intensity responses, LLMs may act as digital "stabilizers" that help reduce conflict escalation and promote more rational dialogue in emotionally charged environments.

From a broader perspective, this pattern enriches our understanding of emotional expression in AI-driven human-computer interaction, particularly in the domain of contentious social issues. The consistent emotional neutrality observed in LLMs challenges existing communication frameworks—such as emotional contagion and social presence theories—which often assume reciprocal emotional engagement between communicators
Together, these two core findings reveal a potential and valuable characteristic of LLM when participating in controversial topic discussions: an inherent tendency to "emotionally debias" or "emotionally cool down".

Practically, this emotional neutrality could be harnessed by platform designers or moderators seeking to mitigate online hostility. For example, LLMs could be integrated into discussion threads to provide balanced, emotion-light perspectives during periods of emotional escalation. However, this potential benefit must be weighed against a crucial caveat: users may interpret persistent neutrality as emotional detachment, avoidance, or even insincerity—especially in contexts where empathy or value signaling is expected. Thus, while LLMs may offer tools for enhancing civility, their impact on user experience and engagement must be carefully evaluated. Over-reliance on emotionally neutral AI responses could unintentionally suppress necessary expressions of concern, passion, or moral stance in public discourse.

\section{Limitation and Future work}
This study enhances understanding of emotional regulation in human-AI interactions while highlighting opportunities for future research to address current limitations.
First, our experimental design focused on Reddit and Twitter datasets. It is worth noting that platforms like YouTube, Instagram, and TikTok demonstrate differing user engagement styles and content ecosystems \cite{Voorveld02012018}.

Secondly, This study may be subject to potential data contamination due to reliance on publicly available content from Twitter and Reddit. As both platforms are openly accessible, portions of their data might overlap with the pretraining corpora of LLMs. As a result, model outputs may reflect memorized patterns rather than authentic reasoning, limiting the generalizability of our findings. Although the extent of such contamination is difficult to measure precisely, we recognize its influence and recommend that future work explore mitigation strategies—such as utilizing post-training datasets that fall outside known LLM training ranges or focusing on content published after established pretraining cutoff dates. Additionally, social media datasets may contain automated content or interactions involving social bots, which could introduce biases. Future studies should implement more rigorous data validation procedures to enhance dataset reliability and minimize potential confounds

Third, we selected five types of models for our study, including commercial and open source ones. The research results may be affected by the specific version of the LLM model and the training data selected. Its universality needs further verification. In the future, we may consider adding more versions of the model for experiments.

Fourth, this study mainly focuses on the emotion analysis of LLM generated texts. Due to experimental conditions, it is not possible to directly examine the actual perception and subjective evaluation of real users on LLM neutral replies. This study mainly analyzes the first round of LLM replies, and does not involve multiple rounds of complex interactions. It is impossible to draw conclusions about the actual impact of LLM neutral replies on subsequent user interaction behaviors and the overall discussion atmosphere in real, dynamic online discussions.
\section{Conclusion}
This study investigates the emotional tendencies of LLMs in responding to climate change discussions on social media. Through cross-platform comparisons and multi-model analysis, we find that LLMs predominantly generate emotionally neutral and low-intensity responses, in contrast to the often polarized and emotionally charged tone of human users. These findings suggest that LLMs may play a moderating role in online discourse, potentially reducing affective escalation and fostering a more balanced communicative environment.

From a social science perspective, this emotional neutrality highlights the emerging role of AI agents as not only communicators but also potential facilitators in public deliberation. However, the implications of such neutrality—whether perceived as constructive moderation or emotional detachment—remain context-dependent and warrant further exploration through user-centered studies and real-world deployment.

\vspace{12pt}


\end{document}